\documentclass[prd,aps,superscriptaddress,showpacs]{revtex4}
\usepackage{epsfig}
\def\beq{\begin{equation}}
\def\eeq{\end{equation}}
\def\bea{\begin{eqnarray}}
\def\eea{\end{eqnarray}}
\def\<{\langle}
\def\>{\rangle}

\usepackage{graphicx}
\usepackage[dvips]{color}
\usepackage{colordvi}
\usepackage{amsfonts}
\usepackage{amssymb}
\usepackage{amsmath}
\usepackage{graphics}
\usepackage{pstricks}
\usepackage{bm}
\usepackage{natbib}

\begin{document}
\title{Transport parameters in neutron stars from in-medium NN cross
sections}
\author{H.F. Zhang}
\affiliation{School of Nuclear Science and Technology, Lanzhou
University, Lanzhou 730000, People's Republic of China}
\author{U. Lombardo}
\affiliation{Dipartimento di Fisica and INFN-LNS, Via S. Sofia 64,
I-95123 Catania, Italy}
\author{W. Zuo}
\affiliation{Institute of Modern Physics, Chinese Academy of
Sciences, Lanzhou 730000, People's Republic of China}

\date{\today}
\begin{abstract}
We present a numerical study of shear viscosity and thermal
conductivity of symmetric nuclear matter, pure neutron matter and
$\beta$-stable nuclear matter, in the framework of the Brueckner
theory. The calculation of in-medium cross sections and nucleon
effective masses is performed with a consistent two and three body
interaction. The investigation covers a wide baryon density range as
requested in the applications to neutron stars. The results for the
transport coefficients in $\beta$-stable nuclear matter are used to
make preliminary predictions on the damping time scales of non
radial modes in neutron stars.
\end{abstract}

\maketitle

\section{Introduction}

Neutron stars are a unique laboratory for studying the equation of
state of nuclear matter at high density and isospin asymmetry,
beyond the ranges typical of heavy ion collisions and nuclei far
from stability. But the observation data are still far from uniquely
constraining the properties of exotic states of nuclear matter. The
detection of gravitational waves could push forward the neutron star
physics. As the gravitational waves  drive the instability of
neutron star oscillations, including non radial modes, the possible
damping mechanisms have to be investigated to justify the existence
of rapidly rotating stars . Good candidates are the viscosity and
the thermal conductivity of the neutron star constituents. The
important role that these parameters can play is the main reason for
the uninterrupted interest in the transport theory of dense matter
over the last three decades \cite{ito}. Microscopic models of
nuclear matter, i.e. based on bare interactions, have been faced
with the interpretation of the equilibrium properties of neutron
stars, such as their structure\cite{exoct} and the onset of
superfluidity \cite{schul}. The same theoretical models have been
also applied to calculate effective mass \cite{effm}and medium
renormalized nucleon-nucleon (NN) cross sections\cite{cross},both
extensively used in the transport-model simulations of heavy ion
collisions. But only recently these two quantities were redirected
to describe the transport properties of nuclear matter. These
quantities are in fact the main ingredients to calculate the
viscosity and thermal conductivity coefficients in neutron stars.
The extension of such calculations to regimes of high density and
isospin, needed in the study of neutron matter core, together with
the interpretation of their inner structure, is a main challenge for
the current microscopic theories of the nuclear matter.

So far all calculations based on microscopic many body approaches
with realistic interactions have shown that the medium effects
result in a noticeable suppression of the NN cross sections
$\sigma_{NN}$\cite{cross} and, as a consequence, in an enhancement
of both  shear viscosity and thermal conductivity. Calculations in
dense nuclear matter have been performed in the T-matrix approach
\cite{sedra}, in the correlated basis function (CBF)
approach\cite{benh1}, and Brueckner-Hartree-Fock (BHF)
approximation\cite{benh2}. Calculations of the transport
coefficients in $\beta$-equilibrium nuclear matter have been
performed with the equation of state (EoS) of asymmetric nuclear
matter from the variational approach \cite{yako}, and from CBF
approach \cite{benh3}. Despite the overall agreement on the medium
effects, different approaches could give different predictions for
transport parameters in $\beta$-equilibrium nuclear matter since
they differ from each other in the different isospin dependence that
becomes more and more visible in higher density range. The results
should be insensitive to the choice of the two body potential
because all realistic potentials used in the microscopic
calculations are accommodated on the experimental NN scattering
phase shifts. But a dependence is expected on the three body force,
especially at the high density, where its influence on the EoS is
dominant. Therefore, it seems interesting to compute the shear
viscosity and the thermal conductivity of nuclear matter, especially
$\beta$-stable nuclear matter, in a wide range of densities needed
for the study of the neutron star core. This will be done within the
BHF approximation. The latter embodies, within a unified
meson-exchange model framework, two and three body forces. The meson
parameters of the two-body realistic interaction (the Bonn B
\cite{BonnB} in our case), which fit the experimental NN scattering
phase shifts in the vacuum, are also adopted to describe the
three-body force \cite{zhl}.
 In this paper the numerical results will be tested, within the simple model
 of constant density neutron star, on the calculation of the dissipation time
 scales to be compared with the time scale of emitting gravitational
 radiations.


\section{Transport parameters}

The transport parameters of Fermi liquids were derived by Abrikosov
and Khalatnikov (AK) from the Landau kinetic equations for a
multicomponent system\cite{ak}
\begin{equation} \label{kin}
\frac{\partial f_i}{\partial t} +
\{f_i,\epsilon_p\}\,=\,\sum_k{\cal\ I}_{ik},
\end{equation}
where $f_i(\vec r,\vec p)$ is the quasiparticle distribution of the
component i , $\epsilon_p$ the quasiparticle energy, and ${\cal\
I}_{ik}$ the collision integral between particles of the components
i and k. From the linearization of the kinetic equations, the shear
viscosity $\eta$ and thermal conductivity $\kappa$ can be extracted
in the AK approximation. The exact expressions, obtained by Brooker
and Sykes after revisiting the resolution of the kinetic
equations\cite{sb1,sb2}, are written as follows
\begin{eqnarray} \label{eta_AK}
\eta T^{2} &=& \frac{1}{20} \rho v^2_F W(\rho)
C(\lambda)\\
\kappa T &=& \frac{1}{12} v^2_F p_F W(\rho) H(\mu),
\end{eqnarray}
where
\begin{eqnarray} \label{probability}
W^{-1}(\rho)&=&\frac{1}{2\epsilon_F}\int_0^{4\epsilon_F}dE\int_0^{2\pi}
\frac{d\theta}{2\pi}\frac{1}{\sqrt{1-E/4\epsilon_F}}\sigma(E,\theta),
\end{eqnarray}
where $\rho$ is the density, $v_F=p_F/m^*$ is the Fermi velocity and
$m^*$ the effective mass. $C(\lambda)$ and $H(\mu)$ are correction
factors corresponding to the exact solution (see for details
\cite{sb1,sb2}). The key quantity is the in medium cross section
here expressed in terms of the energy $E$ in the laboratory frame
and the scattering angle $\theta$ in the center of mass frame. The
upper limit of the energy integration in the average cross section
is four times the Fermi energy of free Fermi gas
$\epsilon_F=p_F^2/2m$ due to the approximation of restricting the
nucleon excitations around the Fermi surface. As described below,
the in medium cross sections are calculated within the Brueckner
theory.
%

\section{In medium cross sections from the Brueckner theory}

In the interior of a neutron star the hadron density can reach
values several time the nuclear matter saturation density, so that
the nucleon-nucleon (NN) collisions are expected to be deeply
affected by the surrounding medium and the corresponding cross
sections can be quite different from those in free space. There are
two main medium effects: first, the NN scattering amplitude is
dominated by the S-wave components of the effective interaction so
that a flattening of the angular distribution is expected in
comparison with the free cross section, which, in the center of mass
frame, is peaked in the forward and backward directions ; second,
the level density in the entrance and exit channels gets reduced by
the strong medium renormalization of the effective mass.

Both effects can be well described in the framework of the
selfconsistent Brueckner theory. In the last years the latter has
made a remarkable step forward by means of the three-body force,
introduced not only to reproduce the empirical saturation properties
of nuclear matter but also to extend to high density the
calculations. The Brueckner theory with two and three-body force is
described elsewhere~\cite{grange,zuo}. Here we simply give a brief
review of the BHF approximation, adopted for the present
calculations. The starting point is the reaction $G$-matrix, which
satisfies the Brueckner-Bethe-Goldstone (BBG) equation,
\begin{equation}
\label{bbg}
G(\omega)=\upsilon_{\mathrm{NN}} +\upsilon_{\mathrm{NN}}
\sum_{k_{1}k_{2}}\frac{ |k_{1}k_{2}\rangle Q_{k_{1},k_{2}}\langle
k_{1}k_{2}|}{\omega -\epsilon_{k_{1}}-\epsilon _{k_{2}}}G(\omega),
\end{equation}
where $k_i\equiv(\vec k_i,\sigma_i,\tau_i)$, denotes the
single-particle (s.p.) momentum, the $z$-component of spin and
isospin, respectively, and $\omega$ is the starting energy. The
$G$-matrix, the Pauli operator $Q$ and the s.p. energies
$\epsilon_k=k^2/2m+U_k$ depend separately on the neutron and proton
densities. The interaction $\upsilon_{\mathrm{NN}}$ given by
\begin{equation}
\upsilon_{\mathrm{NN}} = V^{\mathrm{bare}}_2 + V^{\mathrm{eff}}_3,
\end{equation}
where  $V^{\mathrm{bare}}_{2}$ is the bare two-body force and
$V^{\mathrm{eff}}_{3}$ is the three-body force  averaged on the
third particle as follows
\begin{equation}
\begin{array}{lll}
 \langle\vec r_1 \vec r_2|V^{\mathrm{eff}}_3(T)|\vec r_1^{\ \prime} \vec r_2^{\ \prime} \rangle =
\\[2mm]\displaystyle
 \frac{1}{4}
{\rm Tr}\sum_n \int {\rm d} {\vec r_3} {\rm d} {\vec r_3^{\
\prime}}\phi^*_n(\vec r_3^{\ \prime}) [1-\eta(r_{13}')]
[1-\eta(r_{23}')] \\[6mm]
\times \displaystyle W_3(\vec r_1^{\ \prime}\vec r_2^{\ \prime}
\vec r_3^{\ \prime}|\vec r_1 \vec r_2 \vec r_3) \phi_n(r_3)
[1-\eta(r_{13})][1-\eta(r_{23})].
\end{array}
\label{eq:TBF}
\end{equation}

Since the defect function $\eta(r)$ ($1-\eta(r)$ is the correlated
two body wave function) is directly determined by the solution of
the BBG equation ~\cite{grange}, $V^{\mathrm{eff}}_{3}$ must be
calculated self-consistently with the $G$-matrix and the s.p.\
potential $U_k$ on the basis of the self-consistent BBG equations.
It is clear from Eq.~(\ref{eq:TBF}) that the effective force rising
from the 3BF in nuclear medium is density  dependent via the defect
function. A detailed description and justification of the method can
be found in Refs.~\cite{grange,zuo}.

\begin{figure}[htb]
\centerline {\epsfig{figure=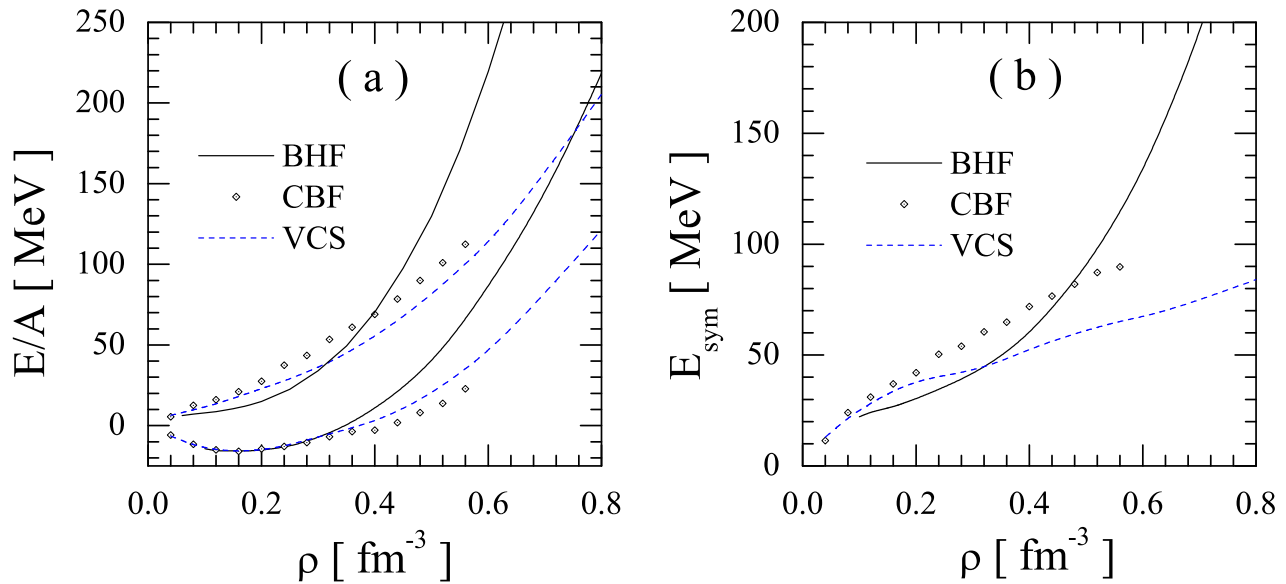,angle=000,width=17cm}}
\caption{(color online).(a) Energy per particle in pure neutron
matter and symmetric nuclear matter.(b) Symmetry energy, calculated
as the difference between the neutron the symmetric nuclear matter
energies. The diamonds represent the results obtained using the
correlated-base function (CBF) approximation with a Fermi gas states
\cite{benh1}, the dashed lines represent the results obtained using
the variational chain summation (VCS)\cite{APR}.\label{eos} }
\end{figure}

In the present calculations the Bonn B potential was adopted as
$V^{\mathrm{bare}}_{2}$~\cite{BonnB}. Besides being a realistic
interaction fitting the experimental the NN scattering phase shifts,
it has the advantage of being built up in terms of meson exchange so
as the three body force. Therefore the choice of the same meson
parameters, that is masses, coupling constants and cutoffs, provides
a unified treatment of two and three-body forces as mentioned
before. Details and results with this interaction are presented in
Ref.~\cite{zhl}. In Fig.\ref{eos}a the equations of state of pure
neutron matter (PNM) and symmetric nuclear matter (SNM) are plotted.
The EoS from the correlated basis theory (CBT)\cite{benh1} and
variational chain summation (VCS) approach\cite{APR} are also
plotted for a comparison , that will be useful for the next
discussion of the transport parameters. A strong deviation from the
Brueckner calculation can be easily observed at increasing density,
which is much more sizeable for the symmetry energy plotted in
Fig.\ref{eos}b. The symmetry energy affects the neutron star
transport properties, because it essentially determines the isospin
composition of the core.

\begin{figure}[htb]
\centerline {\epsfig{figure=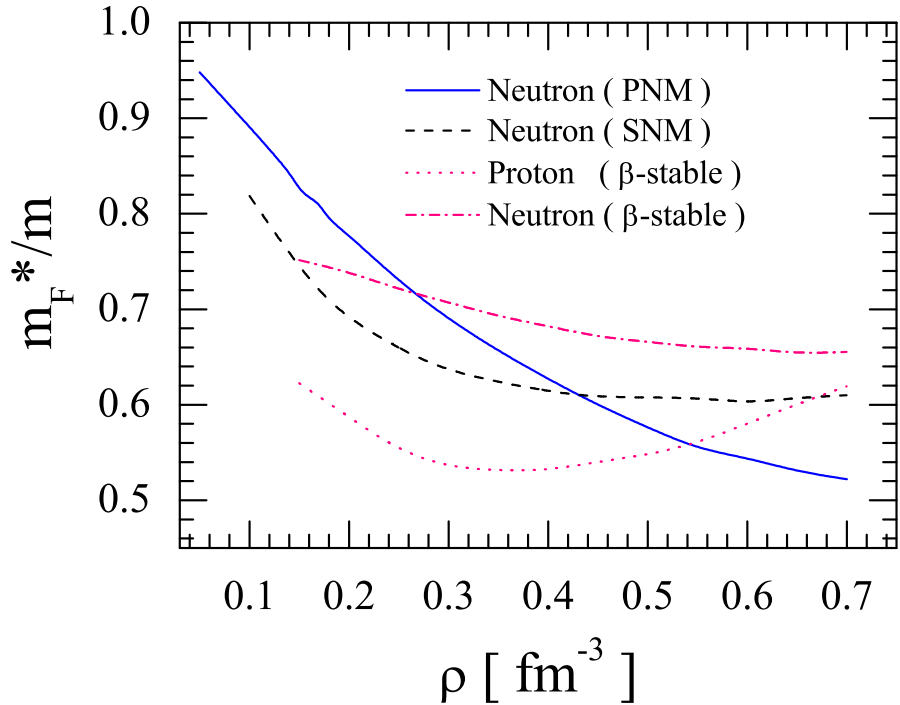,angle=000,width=12 cm}}
\caption{(color online). Density dependence of effective mass
$m_{F}^{*}/m$ in pure neutron matter (blue solid line), symmetric
nuclear matter (black dashed line) and the $\beta$-stable nuclear
matter (pink dotted line), respectively. \label{effmass} }
\end{figure}

In the Brueckner theory  the in-medium NN cross section is  obtained
replacing the T- matrix with the G-matrix and the in vacuum level
density with the in-medium one. This definition is supported by the
property of the G-matrix to go over into the T-matrix in the zero
density limit, as it can be seen from BBG equation (
Eq.(\ref{bbg})).

\begin{figure}[htb]
\label{nn-np} \centering
\includegraphics[angle=0,width=8 cm]{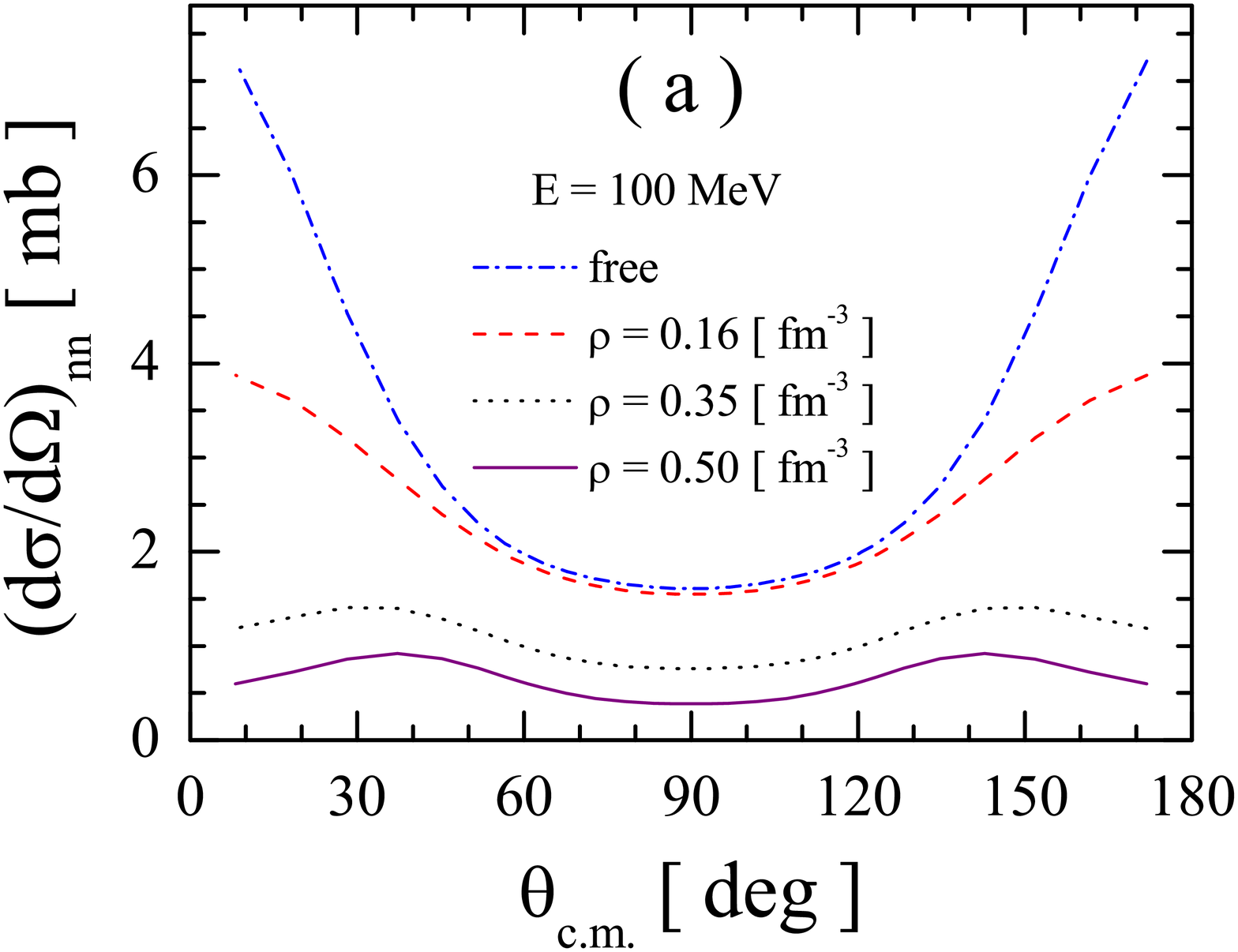}
\includegraphics[angle=0,width=8 cm]{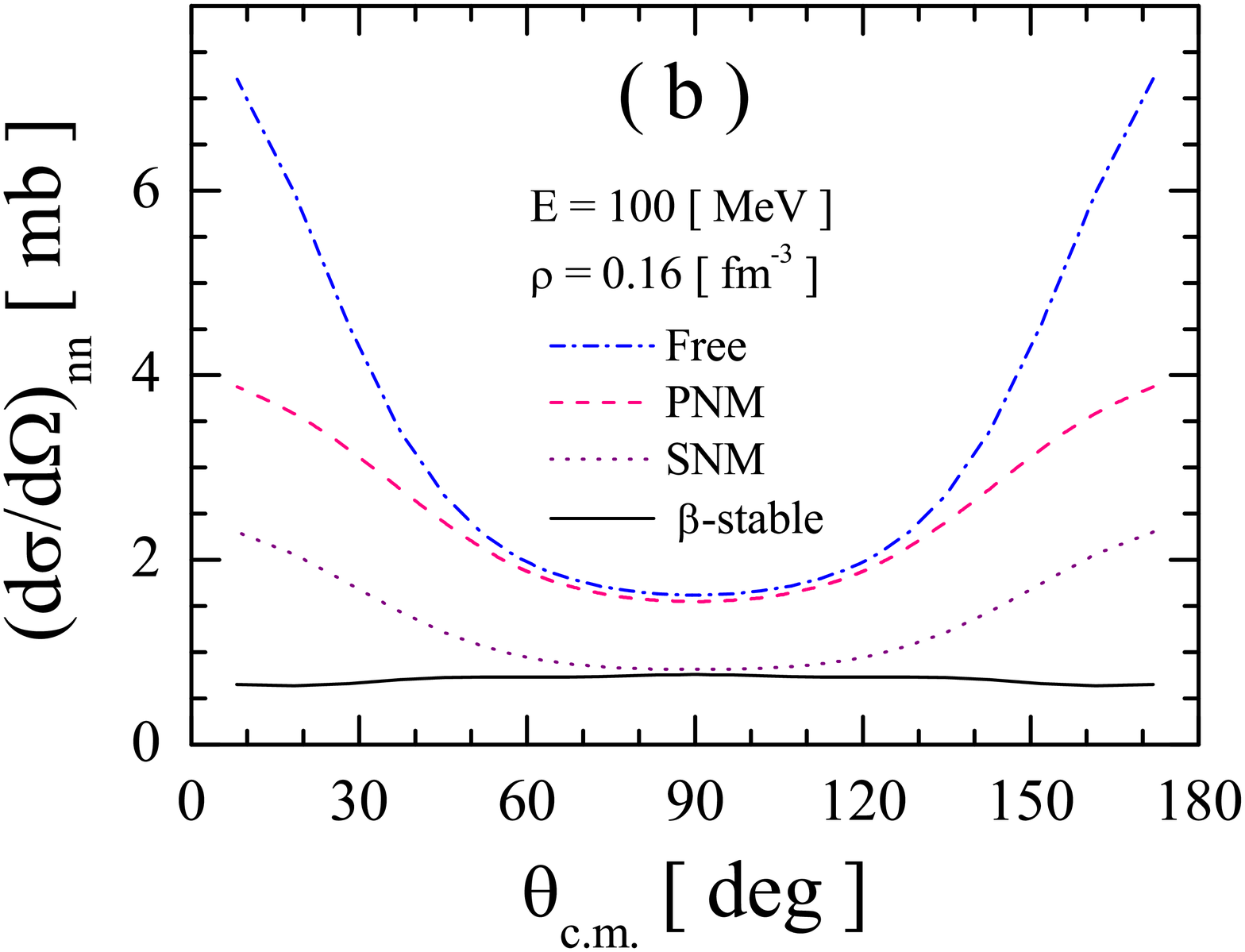}
\includegraphics[angle=0,width=8 cm]{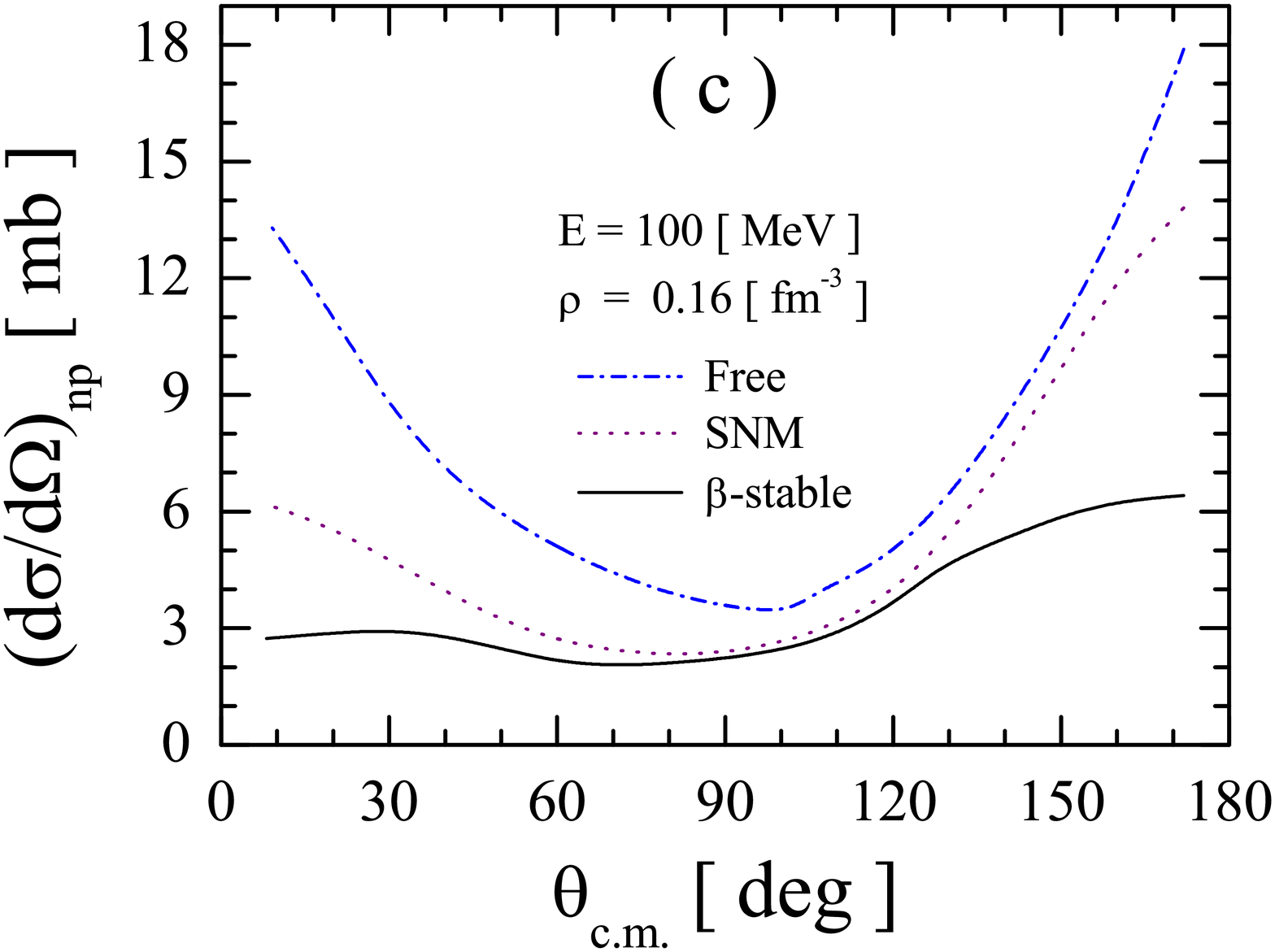}
\caption{(color online). (a) Neutron-neutron differential cross
sections in pure neutron matter. (b) Neutron-neutron  differential
cross section in pure neutron matter, symmetric and $\beta$-stable
nuclear matter. (c) Neutron-proton differential cross section in
pure neutron matter, symmetric and $\beta$-stable nuclear matter.
The free cross section is also plotted for comparison.
\label{cross}}
\end{figure}

In the case of pure neutron matter the neutron-neutron cross section
in the center-of-mass frame is written
\begin{eqnarray}\label{diff1}
\sigma_{nn}(E,\theta)=
\frac{m^{*2}}{16\pi^{2}\hbar^{4}}\sum_{SS_{z}S_{z}^{\prime}}|G_{S_{z}S_{z}^{\prime}}^{S}
(\theta)+(-1)^{S}G_{S_{z}S_{z}^{\prime}}^{S}(\pi-\theta)|^{2}\cdot
\end{eqnarray}
The prefactor is the square of the level density at the Fermi
energy, where the AK approximation assumes the particle transitions
to mainly occur. The medium renormalization of the nucleon mass to
values much less than the unit, as shown in Fig.\ref{effmass},
reduces the level density with respect to the free Fermi gas value.
As a consequence also the in-medium cross section turns out to get
reduced. The additional medium effect, which is incorporated in the
G-matrix, is due to the Pauli blocking which prevents the particles
to scatter into occupied states. In Fig. \ref{cross} (panel (a)) the
in-medium neutron-neutron cross section at the laboratory energy
$E~=~ 100$ MeV, obtained from Eq.(\ref{diff1}), is compared to the
corresponding free one. As mentioned above, the differential cross
section becomes more and more isotropic at increasing density. In
addition, its magnitude is reduced, that is a common feature  to all
predictions based on microscopic approaches. In all curves the
angular distribution is symmetric around $90$ degrees, since the
cross section section is antisymmetrized for identical particles. In
the panel (b) the neutron-neutron cross section in various baryon
environments is reported. It is worth noticing that it is completely
isotropic in $\beta$-stable matter.

In nuclear matter, besides the neutron-neutron scattering,
neutron-proton scattering must be considered. In this case the cross
section is
\begin{eqnarray}\label{diff2}
\sigma_{np}(E,\theta)=
\frac{m^{*2}}{16\pi^{2}\hbar^{4}}\sum_{SS_{z}S_{z}^{\prime}}
|G_{S_{z}S_{z}^{\prime}}^{S}(\theta)|^2 \cdot
\end{eqnarray}
In Fig.\ref{cross}c the corresponding cross sections are depicted
for symmetric nuclear matter and $\beta$-stable matter. A common
feature to all cases is the medium suppression, but a difference is
to be noticed in the angular distribution. The enhancement in the
backward direction of $\sigma_{np}$ is a signature of the
anisotropic behavior of the tensor force in the SD channel, giving
the dominant contribution to the neutron-proton interaction
\cite{machleidt, Lombardo}.

In the core of neutron stars the neutron and proton composition is
determined by the condition of equilibrium with leptons (electrons
and muons), assuming total charge neutrality. Thus for a given total
baryonic density the proton and neutron fractions are determined by
the chemical equilibrium condition
\begin{equation}
\label{fractions} \mu_n-\mu_p\,~=~\,4\beta E_{sym}\,~=~\mu_e
\end{equation}
where $\mu_n$, $\mu_p$ and $\mu_e$ are the chemical potential of
neutrons, protons and electrons (no muons for simplicity),
respectively. The electron chemical potential is determined by the
charge neutrality with protons, assuming the electrons to form a
free Fermi gas. The crucial property is the density dependence of
the symmetry energy which determines the imbalance between proton
and neutron fractions.
 In general the nuclear system in such state is
strongly isospin asymmetric. As a consequence, the calculations of
the transport coefficients must be extended to asymmetric nuclear
matter. Eq.(\ref{diff2}) for the neutron-proton cross section is to
be replaced by the following one
\begin{eqnarray}
\sigma_{np}(E,\theta)= \frac{1}{16\pi^{2}\hbar^{4}}(\frac
{2m^*_nm^*_p}{m^*_n+m^*_p})^2\sum_{SS_{z}S_{z}^{\prime}}
|G_{S_{z}S_{z}^{\prime}}^{S}(\theta)|^2
\end{eqnarray}
For small asymmetries the isospin effect on the 'reduced' effective
mass is of the $\beta^2$ order, because $m^*_n\approx m^*_0+\beta m$
and $m^*_p\approx m^*_0-\beta m$, where $m^*_0$ is the effective
mass \cite{machleidt, Lombardo} in symmetric nuclear matter.

\section{Numerical results and comparison with other calculations}

\begin{figure}[htb]
\centering
\includegraphics[angle=0,width=5.2 cm]{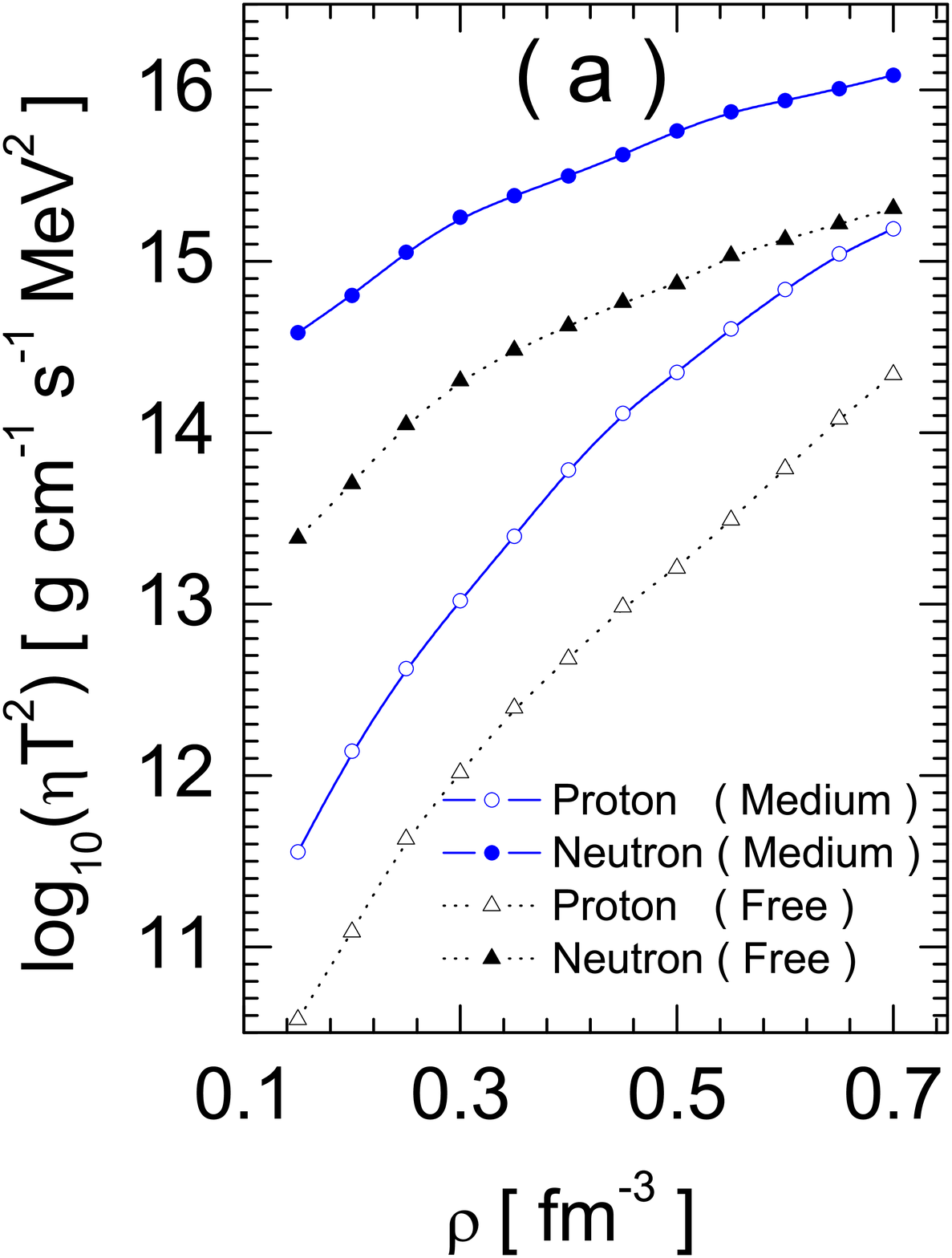}
\includegraphics[angle=0,width=5.5 cm]{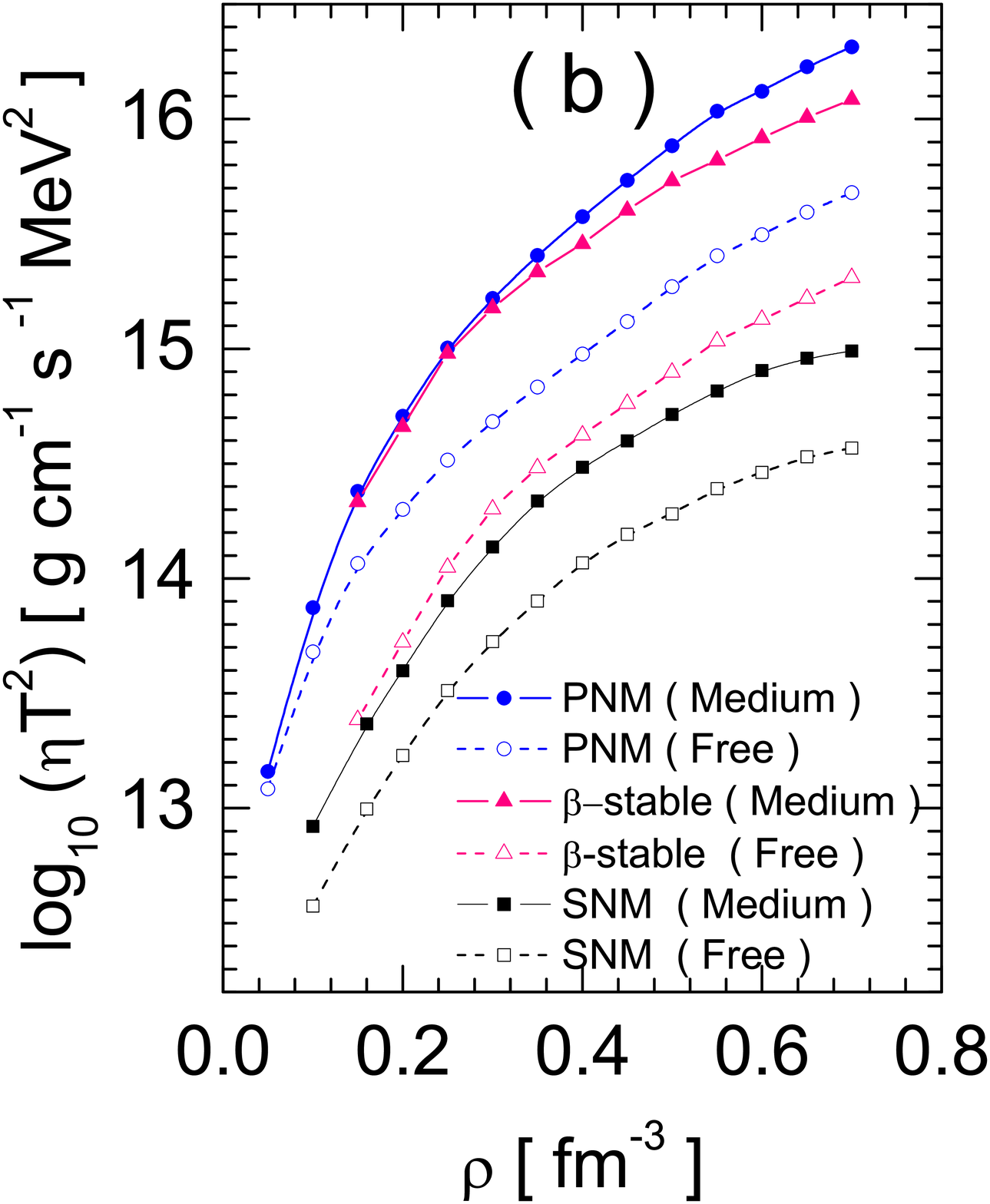}
\includegraphics[angle=0,width=5.5 cm]{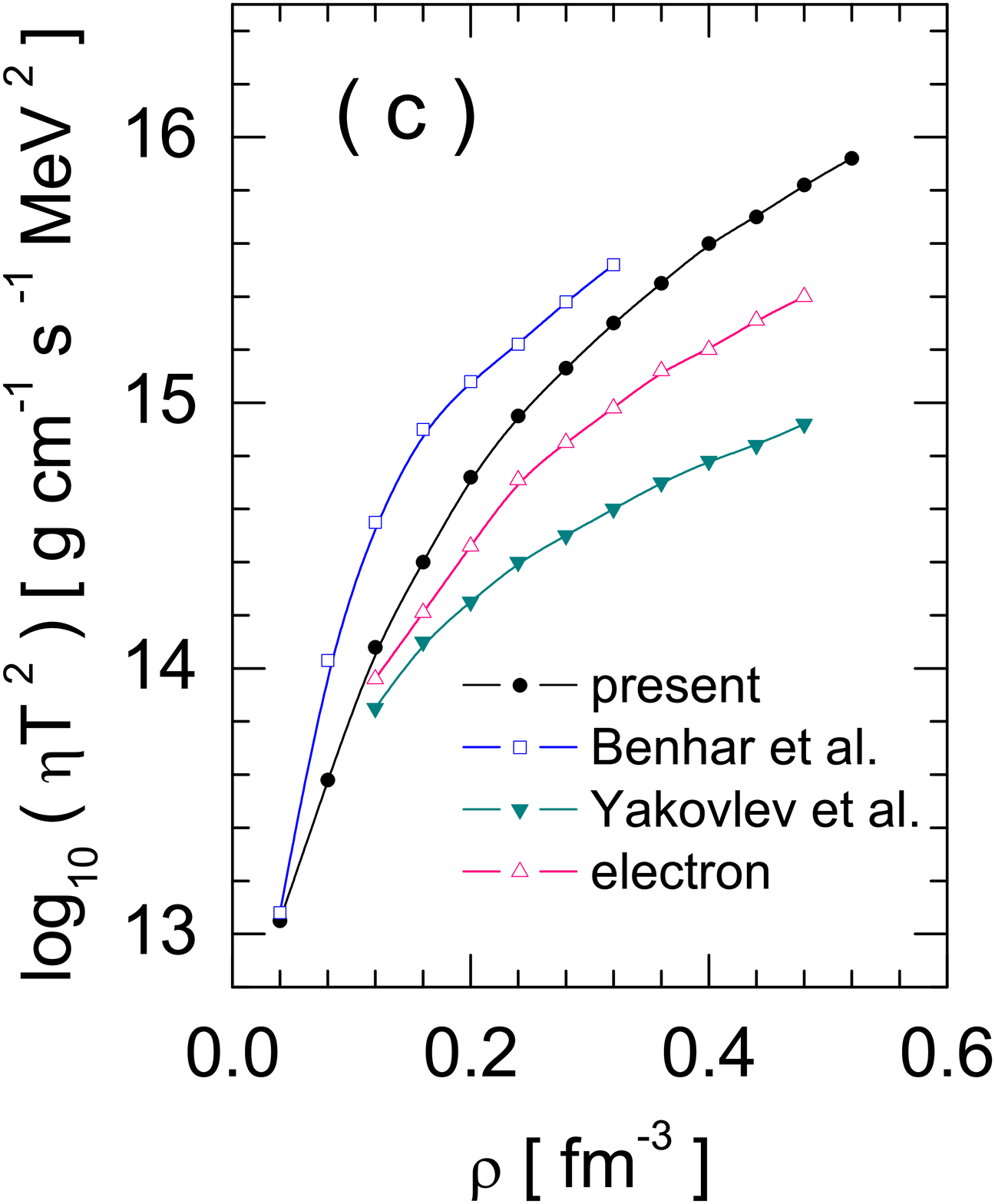}
\caption{(color online).  Shear viscosity from the Brueckner theory.
(a) Neutron and proton shear viscosity in $\beta$-stable nuclear
matter. (b) Neutron viscosity in all considered states. (c)
Comparison of the Brueckner neutron viscosity with other
calculations (see text) and with the electron viscosity, in
$\beta$-stable nuclear matter.} \label{viscosity}
\end{figure}

The shear viscosity $\eta$ vs. density was calculated for the three
different nuclear matter configurations above considered, i.e.
neutron matter, symmetric nuclear matter, and $\beta$-stable
asymmetric nuclear matter. In Fig.\ref{viscosity}a the neutron and
proton viscosities from free and in-medium cross sections  are
plotted vs. density in the case of $\beta$-stable nuclear matter. At
low total density the difference between neutrons and protons is
about three orders of magnitude because the proton fraction is quite
small, but it reduces to only one order of magnitude at high density
where the proton fraction becomes $30\%$ of the total density. As
expected, both viscosities, corresponding to free and in-medium
cross section, are increasing with density, and the medium
enhancement turns out to be  quite large indeed. In
Fig.\ref{viscosity}b the neutron viscosity in pure neutron,
symmetric nuclear and $\beta$-stable matter, is plotted and the
medium effects are also emphasized by the comparison with the free
case. The medium effect is a strong enhancement of $\eta$ ,mainly as
a consequence of the reduction of the level density, more pronounced
for $\beta$-stable matter, less for symmetric nuclear matter.
Comparing neutron matter and $\beta$-stable matter, the values are
very close to each other in the low density range, where
$\beta$-stable matter is essentially made of neutrons, then they
deviate for the increasing weight of the proton fraction, according
to the $\beta$-stability condition (see Fig.\ref{cross} and
Eq.(10)).

Fig.\ref{viscosity}c shows the comparison among various microscopic
calculations. The data from Benhar et al.\cite{benh3}, available up
to a maximum range $0.3 fm^{-3}$, differ from the present
calculation for the different neutron-to-proton composition of
$\beta$-stable nuclear matter. In that case the proton fraction, at
the same total density, is larger, as a consequence of a larger
symmetry energy (curve CBF in Fig.\ref{eos}), and thus the influence
on the neutron viscosity by the neutron-proton cross section turns
out to be bigger. The data from Yakovlev et al.\cite{yako} were
obtained from the constant effective mass approximation ($m^*/m =
0.8$ in the plot), which is not viable for the  effective mass
gradual quenching at higher density. The result is that the neutron
viscosity is underestimated and it turns out to be smaller than the
electron viscosity. The opposite happens in the present as well as
the Benhar et al. calculation, as shown in Fig.\ref{viscosity}c.
Therefore the electron viscosity would definitely be immaterial in
the study of the energy dissipation.

\begin{figure}[htb]
\centering
\includegraphics[angle=0,width=7. cm]{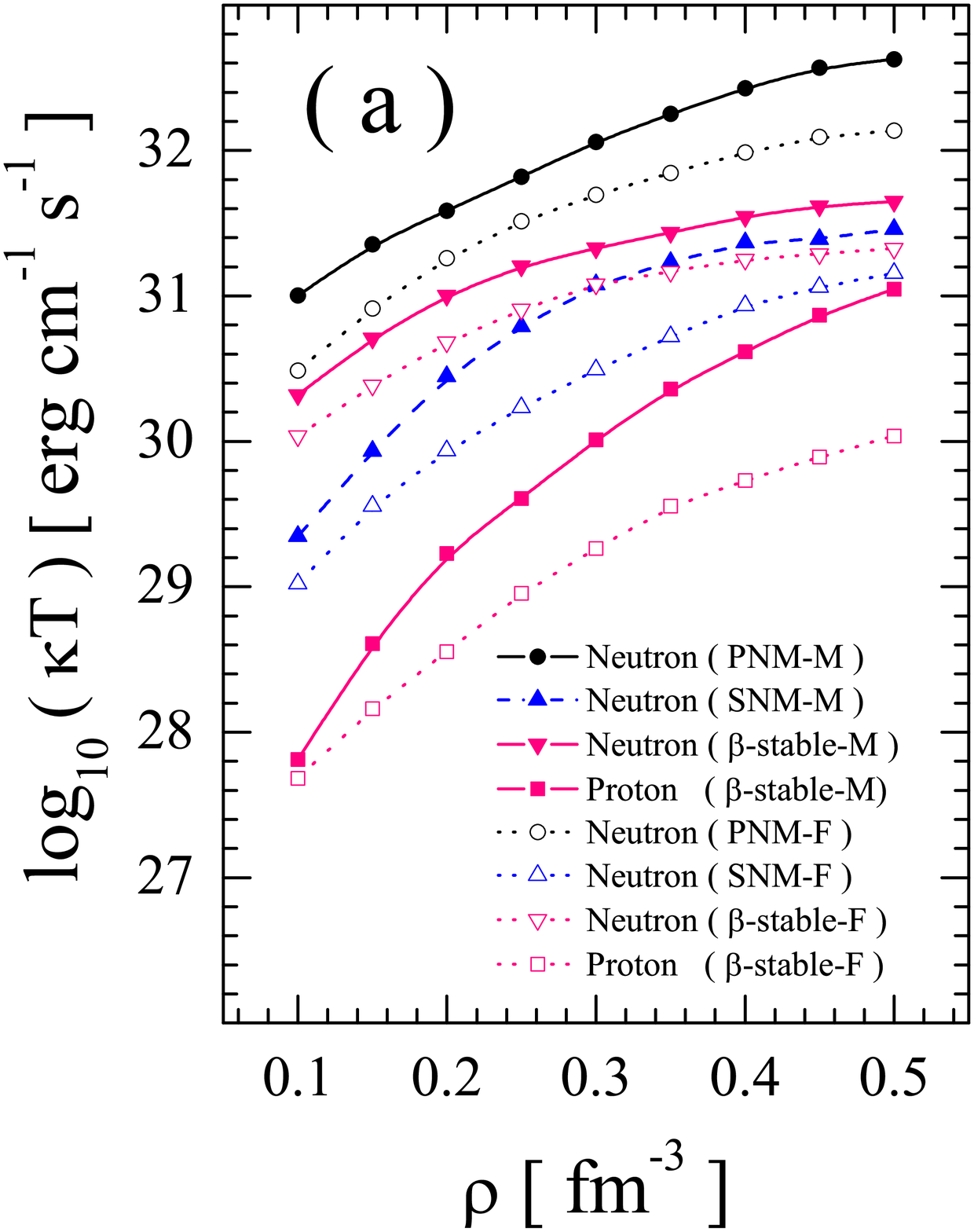}
\includegraphics[angle=0,width=7.2cm]{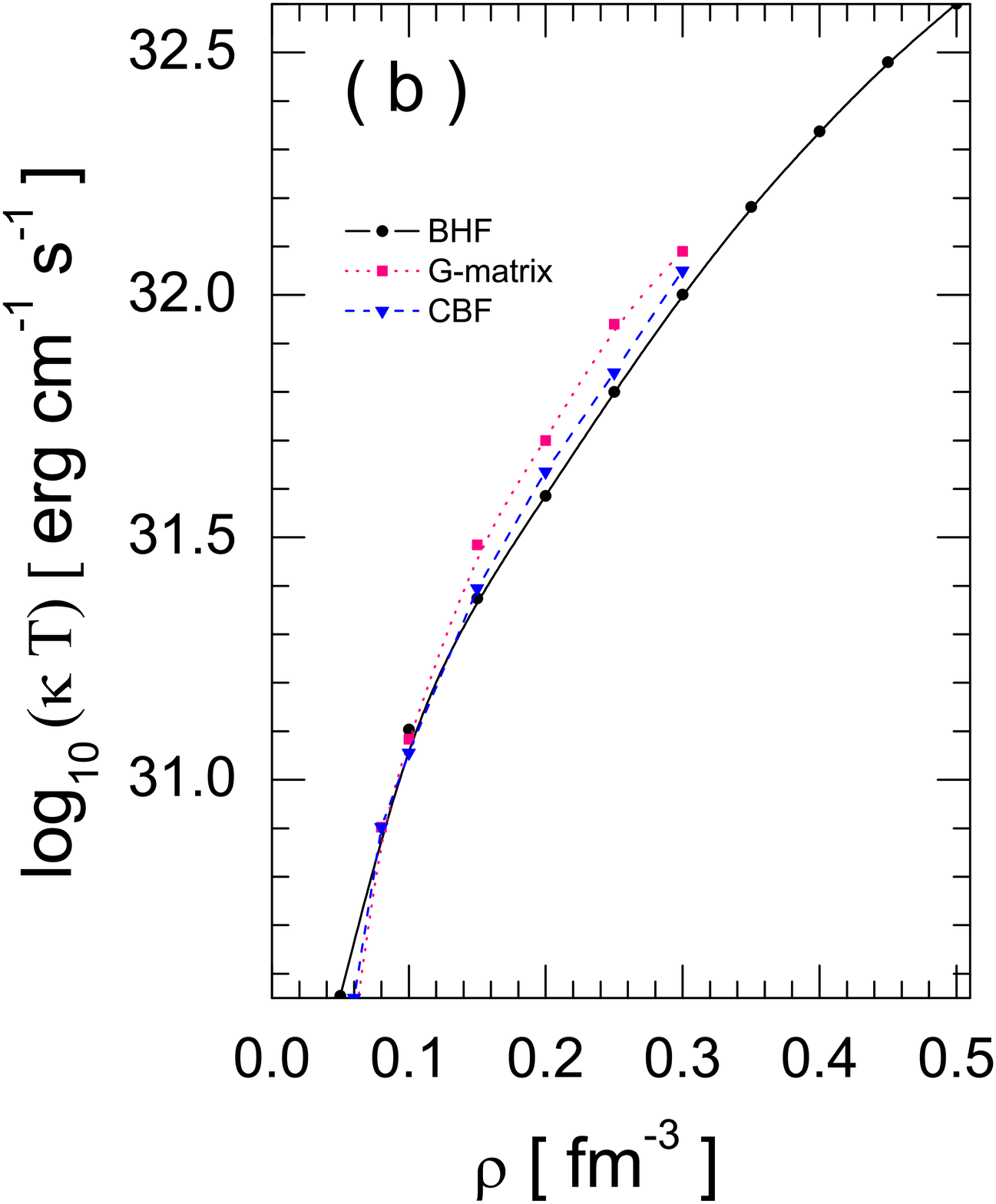}
\caption{(color online). (a) Thermal conductivity from the Brueckner
theory in all considered states. (b) Comparison of the thermal
conductivity in pure neutron matter between the BHF approximation
and two other recent calculations(see text).} \label{cond}
\end{figure}
The thermal conductivity $\kappa$ was also calculated according to
Eq.(3). The results in the various nuclear matter configurations are
reported in Fig.\ref{cond}. Again the medium corrections are
emphasized by plotting the $\kappa$ values obtained with in-vacuum
and in-medium cross sections. In the right panel the present results
are compared with the two above mentioned models \cite{SY,benh1}. To
assess the extent of variations it must be taken into account that
in this case the scale is quite different from that of
Fig.\ref{viscosity}c. In any case the good agreement among the
different models for the pure neutron conductivity confirms that the
deviations can only be traced to the different isospin dependence.


The transport parameters, viscosity and thermal conductivity,
calculated for $\beta$-stable nuclear matter can be used to
determine the respective time scales of energy dissipation $\tau_V$
and $\tau_T$ . This requires to integrate $\eta(\rho)$ and
$\kappa(\rho)$ weighted with the density profile $\rho(r)$ of a
given neutron star configuration. The latter can be obtained solving
the Tolman-Oppenheimer-Volkov (TOV) equation with a given equation
of state of nuclear matter. Simplified expressions for the time
scales governing the energy dissipation from non radial oscillations
can be derived for a quasi-uniform density model \cite{lind}:
\begin{eqnarray} \label{tscale}
\tau_V^{-1}&=&(l-1)(2l+1)\frac{\eta}{\rho R^2}
\\
\tau_T^{-1} &=& 0.0034 \frac{l^3(2l+1)}{l-1}\frac{\kappa T}{G\rho^2
R^4}
\end{eqnarray}
where R is the radius of the star, $\rho$ is the density calculated
as the ratio between the mass M and the volume $4\pi R^3/3$.
\begin{table}[t]
\centering
\includegraphics[angle=0,width=12.cm]{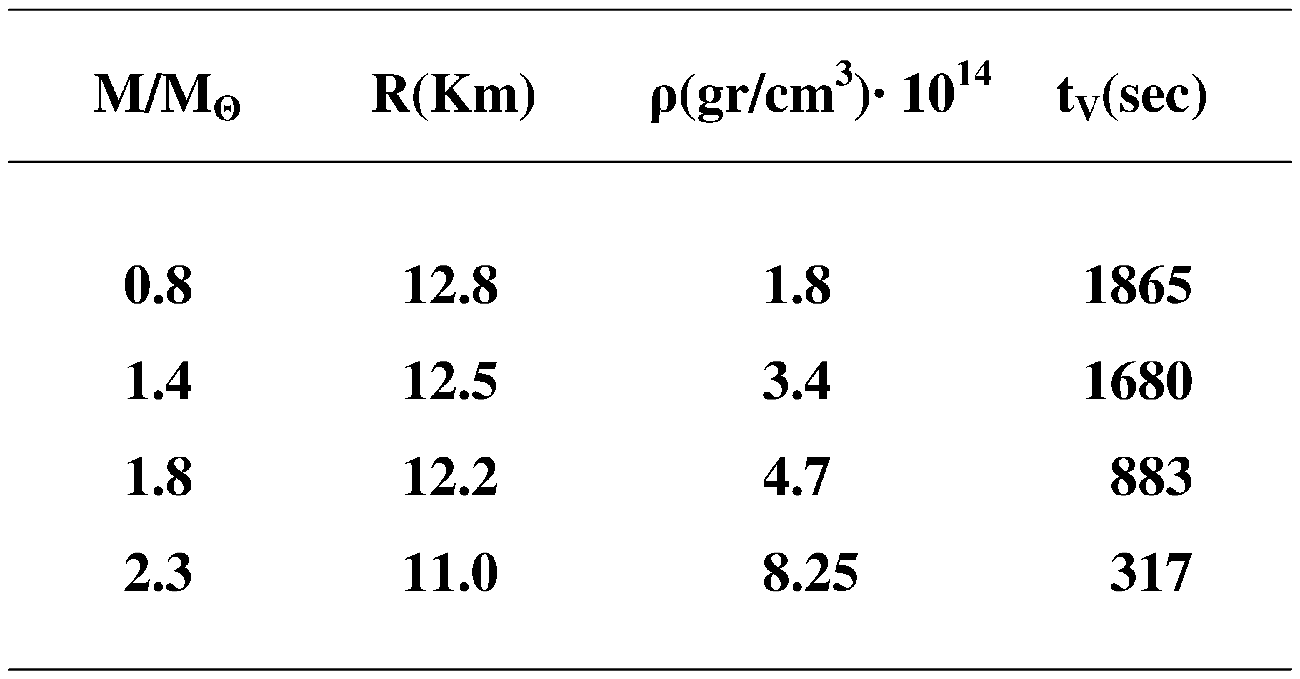}
\caption{Non-linear mode (l=m=2) damping time scale for energy
dissipation due to neutron viscosity in neutron stars.} \label{time}
\end{table}
The parameter $l$ is the angular momentum of the non radial
oscillation $Y_{lm}$ (below we consider l=m=2). As discussed in
Ref.\cite{lind}, Eqs.(12)-(13) underestimate the values for the
transport parameters \cite{lind} within a factor about $10$, because
the constant mass approximation smoothes out their increase at
higher densities. In Table I the damping time scales of $Y_{22}$ non
radial modes are reported for a number of neutron star mass-radius
configurations. The neutron star configurations are taken from
Ref.\cite{hybrid}, where the TOV equation was solved using the
equation of state derived from the same BHF approximation as for the
in-medium cross sections. The low temperature results show that,
within the above mentioned uncertainty of constant mass
approximation, the time scale for energy dissipation due to the
neutron shear viscosity could be of the same order of magnitude as
the time scale associated to gravitational waves $(10-100 sec)$, at
least at high density. The proton contribution is much less
important, being  the proton fraction much less than the total
density. In other words the instability driven by gravitation
radiation could be prevented by the shear viscosity dissipation in
cold neutron stars. To confirm such a statement, a more accurate
calculation is to be performed by means of the viscosity integration
weighted with the neutron star density profiles $\rho(r),0\le r\le
R$. On the contrary, the high temperature time scales are too large
by several orders of magnitude ($\tau_V\approx 10^{13}\div10^{14}
sec$, at $T= 3\cdot10^{11}K$) and no help can be expected by
superfluidity since the critical temperature is much less that the
newborn star temperature.

The energy dissipation due to thermal conductivity $\tau_T$ can be
calculated from Eq.(12) for a non radial mode $l=2$. At temperature
$T=10^6 K$ it is about $10^3 \tau_V$ and it gets increased at higher
temperature. Therefore its effect on the damping can be neglected in
agreement with other calculations\cite{lind}.

\section{Conclusions}

The transport parameters, shear viscosity and thermal conductivity,
of neutron stars have been calculated in the framework of the BHF
approximation with two and three body forces. Both forces are
described, in a unified treatment, by the one-boson exchange model
with the meson parameters of the realistic Bonn B potential. In the
Brueckner theory the in-medium NN cross sections are calculated
replacing  the in vacuum scattering amplitude with the G-matrix and
the nucleon bare mass with the effective mass. The in-medium strong
effective mass renormalization, which affects the level density in
the entrance and exit channels, is mainly responsible  for the
strong deviation of the in-medium NN scattering cross sections from
the scattering in free space. The main result is a remarkable
enhancement of the transport coefficients. This effect is well known
since some decades, but only recent  $\it ab \, initio$ calculations
provide reliable quantitative estimates in domains of nuclear
matter, which do not benefit from direct empirical constraints. The
underlying many body approaches, in fact, are based on realistic NN
interactions without free parameters, and are able to calculate on
the same footing both the equation of state of nuclear matter
determining the  composition of neutron stars as well as the
transport properties parameters determining the neutron star cooling
and the damping of collective motions.

The calculation of the transport parameters was first performed for
pure neutron matter and symmetric nuclear matter, then it was
extended to $\beta$-stable nuclear matter for the sake of
application to neutron stars. The calculation covers a wide density
range, as requested by the study of the neutron-star core. The
numerical results are compared with other recent $\it ab \, initio$
calculations. Concerning the shear viscosity, the BHF prediction is
such that the neutrons give a contribution larger than the
electrons, in contrast to the Yakovlev et al. calculation
\cite{yako}, where the constant effective mass approximation is
adopted. On the other hand, the different density dependence of the
symmetry energy is responsible for the different shear viscosity
obtained by Benhar et al.\cite{benh3} in beta-stable configurations.

A preliminary estimate of time scales of non radial mode damping
shows that only in cold stars the dissipation times due to viscosity
are comparable with the gravitational radiation time, whereas the
effect of thermal conduction is negligible. But these results should
be confirmed by more accurate calculations beyond the constant
density approximation. Such calculations are on the way.

In order to interpret the spin and thermal evolution, the neutron
stars have been assumed to be in a superfluid state, that could
deeply influence the their transport properties as well. But the
recent calculations of the neutron-neutron and proton-proton gaps
indicate that the proton $^1S_0$ pairing is present only in a
restricted region of low density, where the proton fraction in
$\beta$-equilibrium with neutrons is quite small\cite{zuo1}, and the
high density neutron gap from the channel $^3PF_2$ is so small that
it could be easily suppressed by even weak polarization effects
\cite{zuo2}. Despite this theoretical uncertainty, the onset of
superfluidity is supported by the phenomenology, and its role in the
damping of collective modes should be clarified.

For the application of the transport properties to neutron stars,
one should also include strange components in the $\beta$-stable
nuclear matter. In fact the strange particle fractions are
increasing with density in the inner core, where they can contribute
with the 20 per cent to the total baryonic matter\cite{yper}. In the
case of hybrid stars the quark matter can compete with baryons;
therefore the contribution to transport properties of quarks must be
considered with a consistent treatment of hadron-to-quark transition
\cite{hybrid}. The last two research issues, superfluidity and
transport properties of non nucleonic components are presently under
investigation.

\section{Acknowledgements}
H.F.Z is very grateful to the nuclear theory group for valuable
discussions in LNS-INFN, where the work is done. This work is
supported by the Natural Science Foundation of China (Grants
10775061, 10875152, 10875151, 10975064 and 10740420550), by the
Fundamental Research Fund for Physics and Mathematics of Lanzhou
University (Grants LZULL200805), the Fundamental Research Funds
for the Central Universities (grants lzujbky-2009-21), by the CAS
Knowledge Innovation Project NO.KJCX-SYW-N02, The Major State
Basic Research Developing Program of China (2007CB815004), by
European Commission under Grants No. CN/ASIA-LINK/008(94791).


\end{document}